\documentclass[prx,nobibnotes,nofootinbib,twocolumn,showpacs]{revtex4-2}
\usepackage{mathrsfs}
\usepackage{calc}
\usepackage[T1]{fontenc}
\usepackage[latin1]{inputenc}
\usepackage{amsmath}
\usepackage{amssymb}
\usepackage{xspace}
\usepackage{graphicx}
\usepackage[normalem]{ulem}
\usepackage{xcolor}
\usepackage{multirow}

\usepackage{tikz}
\usetikzlibrary{arrows}
\usetikzlibrary{calc,decorations.pathmorphing,shapes}

\allowdisplaybreaks[1]

\newcommand{\be}{\begin{equation}}
\newcommand{\ee}{\end{equation}}

\definecolor{darkgreen}{rgb}{0.2,0.6,0}
\definecolor{lightblue}{rgb}{0,0.5,0.8}
\definecolor{lightred}{rgb}{0.8,0.2,0.2}
\definecolor{darkorange}{rgb}{1,0.549,0}
\definecolor{brown}{rgb}{0.609, 0.164, 0.164}

\makeatother

\begin{document}

\definecolor{rvwvcq}{rgb}{0.08235294117647059,0.396078431372549,0.7529411764705882}
\definecolor{wrwrwr}{rgb}{0.3803921568627451,0.3803921568627451,0.3803921568627451}

\title{Reply to "Comment" on "Regular evaporating black holes with stable cores"}
\author{Alfio Bonanno}
\affiliation{INAF, Osservatorio Astrofisico di Catania, via S.Sofia 78 and INFN, Sezione di Catania, via S. Sofia 64, I-95123,Catania, Italy}
\author{Amir-Pouyan Khosravi}
\author{Frank Saueressig}
	\affiliation{Institute for Mathematics, Astrophysics and Particle Physics (IMAPP), \\ Radboud University, Heyendaalseweg 135, 6525 AJ Nijmegen, The Netherlands}

\begin{abstract}
We reply to the ``Comment'' on ``Regular evaporating black holes with stable cores'' by R.\ Carballo-Rubio, F.\ Di Filippo, S.\ Liberati, C.\ Pacilio, and 
M.\ Visser. As a key result, we show that the regime of mass-inflation identified in the comment connects smoothly to the late-time attractors discovered in our works [A.\ Bonanno \emph{et.\ al.}, Regular black holes with stable cores, Phys.\ Rev.\ D {\bf 103}, 124027 (2021) and Regular evaporating black holes with stable cores, Phys.\ Rev.\ D {\bf 107}, 024005 (2023)]. Hence, the late-time stability of regular black holes is not affected by this intermediate phase.
\end{abstract}
\maketitle
%----------------------------------------
\section{Introduction}
%----------------------------------------
Regular black holes are a phenomenologically interesting alternative to the black hole solutions arising from general relativity. In this context, it is important to clarify whether these alternatives remain regular once small perturbations are included. Extrapolating from the Reissner-Nordstrom solution, the presence of a Cauchy horizon could lead to a dynamical instability of the regular geometry through the mass-inflation effect \cite{Poisson:1989zz,PI1990,Ori:1991zz}. Our works \cite{Bonanno:2020fgp,Bonanno:2022jjp} established that specific classes of regular black hole solutions, including the Hayward geometry \cite{Hayward:2005gi}, do not exhibit an exponential growths of the mass function at asymptotically late times, despite the presence of a Cauchy horizon. The comment \cite{Carballo-Rubio:2022twq} discusses this conclusion in a critical way. In this reply, we clarify the global picture arising from the various claims made in the literature.

%------------------------------------
\subsection{What we have done}
%------------------------------------
 We reported a stability analysis of static, regular black hole geometries \cite{Bonanno:2020fgp} and dynamical regular black hole geometries emitting Hawking radiation \cite{Bonanno:2022jjp}. Following the lines of the Ori model \cite{Ori:1991zz}, we modelled the perturbations by an infinitely thin shell impacting on the Cauchy horizon. As our main result, we established the existence of late-time attractors which tame the instability due to the mass-inflation effect. We also identified the conditions on the regular geometry that give rise to the modified dynamics. The results of our dynamical analysis superseed earlier claims \cite{Carballo-Rubio:2018pmi}, which argued the presence of the mass-inflation instability based on a non-dynamical analysis employing the Dray-t'Hooft-Redmount (DTR) relations \cite{Bar-Isr}. 

\subsection{Why is our analysis trustworthy?}
Our dynamical analysis is based on the Ori-model \cite{Ori:1991zz}. Thus, we work with the same simplifying assumptions that were used when establishing the mass-inflation effect for the Reissner-Nordstrom black hole \cite{Poisson:1989zz}. We carefully cross-checked that we reproduce the instability encountered in this case. Analyzing the same model in a regular black hole background then allows a direct comparison of the dynamics in the vicinity of the Cauchy horizon. This showed beyond any doubt that certain regular black hole geometries, {when perturbed}, do not exhibit an exponential growth of the mass function at late times. Since it is this exponential growth that underlies the notion of mass-inflation, it is fair to say that the mass-inflation effect is absent for these geometries. \\

\subsection{On extending the stability analysis \\ to  early times}
The central claim made in \cite{Carballo-Rubio:2021bpr} and iterated in the comment \cite{Carballo-Rubio:2022twq} is that the mass-inflation effect destabilizes the regular black hole geometry \emph{before} the late-time attractors identified in \cite{Bonanno:2020fgp,Bonanno:2022jjp} are reached.  
To substantiate this assertion, the Ori model is extended into the early-time domain. In this case, it is imperative to pay attention to the initial conditions which need to be imposed sufficiently close to the Cauchy horizon so that the perturbative solution is valid. Otherwise, the extrapolation \emph{lacks reliability}. This is underscored by Poisson and Israel, and collaborators \cite{PI1990, Poisson:1989zz, Anderson:1993ni,Bonanno:1994ma,Bonanno:1994qh}, who emphasize that conclusions drawn from this model rest upon three pivotal assumptions: 1) the applicability of the optical geometric limit for both inward and outward propagating gravitational waves, 2) the permissibility of neglecting the matter flow from the collapsing star and 3) the existence of the Cauchy horizon $r_0$ at $v=\infty$, with $v$ being the advanced time. These simplifying assumptions do not necessarily hold at early times. \\

\begin{figure}[t!]
	\includegraphics[width=0.95\columnwidth]{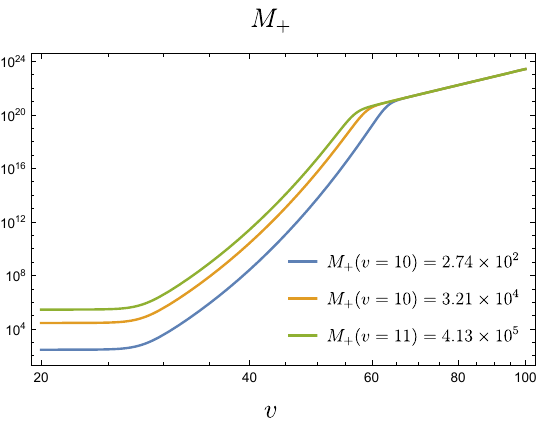}
	\caption{\label{fig.1} Log-log plot of the Misner-Sharp mass $M_+(v)$ obtained from integrating the Ori-model based on the Hayward geometry for several initial conditions. The exponential growths of the mass function at intermediate time-scales ($30 \lesssim v \lesssim 60$) connects smoothly to the late-time attractor quenching the mass-inflation effect at $v > 70$.}
\end{figure} 
\subsection{Connecting the scaling regimes} 
An interesting point raised by the comment \cite{Carballo-Rubio:2022twq} is that there could be an intermediate regime which is characterized by the black hole having shed a sufficient amount of the initial perturbations so that the dynamics can be captured by the Ori model while the dynamics is not yet controlled by the late-time attractor discovered in our works. It is argued that this phase exhibits an exponential growths of the mass function. This may be fatal for two reasons:
\begin{enumerate}
\item The spacetime curvature may reach values where the model is no longer applicable.
\item The mass-function may be driven to values where the dynamics fails to connect to the late-time attractor.
\end{enumerate} 
The first point is not applicable, since the Ori model is actually used to infer the dynamical build-up of a curvature singularity at the Cauchy horizon. On this basis, the model is not used outside the traditional analysis carried out for the Reissner-Nordstrom geometry. 

Clarifying the second point requires a robust numerical analysis of the basin-of-attraction for the late-time attractor. This is currently missing in the literature. On this basis, we revisited the dynamics of the Hayward model with $l = 1/2$ and  initial conditions set at $R(v) - r_{0} = 0.01$. This guarantees that we are close enough to the Cauchy horizon, so that a perturbative analysis is on secure ground. The numerical integration of the dynamics then shows that there are solutions where the exponential increase of the mass function smoothly connects to the late-time attractor. These solutions are shown in Fig.\ \ref{fig.1}. This establishes that the dynamics at intermediate time-scales is by no means fatal for the regular black hole solution. \\

\subsection{Towards realistic models of black holes} 
Our study focuses on an idealized situation where effects related to the formation of the black hole, accretion, or its interaction with the cosmic microwave background are neglected. This is the typical setting used when discussing, e.g., black hole thermodynamics and constitutes a standard assumption in the field. While it is interesting to investigate such effects, this discussion is beyond the scope of our initial works. \\

\section{Conclusion}
The comment \cite{Carballo-Rubio:2022twq} identified a transient phase of exponential growth in the black hole mass function which could potentially destabilize regular black hole geometries. We show that this phase connects smoothly to the late-time attractor discovered in our works \cite{Bonanno:2020fgp,Bonanno:2022jjp}. Hence our conclusions about the late-time stability of regular black hole solutions, also reviewed in \cite{Bambi:2023try}, remain valid.

%As initially indicated in \cite{Bonanno:1994qh} and elaborated further in our contribution to the comprehensive review on regular black holes \cite{Bambi:2023try}, gaining insight into the early-time dynamics of the geometry and dispelling the assumptions concerning the Cauchy horizon's existence necessitate a meticulously detailed dynamical depiction of the collapse process leading to the formation of the regular black hole. Consequently, while the assertions presented in \cite{Carballo-Rubio:2021bpr} and reiterated in the comment \cite{Carballo-Rubio:2022twq} are mathematically correct, they rest on very speculative foundations.


\end{document}